\RequirePackage[mathlines]{lineno}
\documentclass[prd,twocolumn,nofootinbib,showpacs,amsmath,amssymb]{revtex4-2}
\usepackage{overpic,graphicx}
\usepackage{dcolumn}
\usepackage{bm}
\usepackage{rotating}
\usepackage{subfigure}
\usepackage{color}
\usepackage[bookmarksnumbered, pdfstartview=FitH,colorlinks,urlcolor=blue, citecolor=blue,linkcolor=blue,] {hyperref}
\usepackage{lineno}
\usepackage{multirow}
\usepackage{amsmath}
\usepackage{makecell}

\begin{document}


\title{\bf \boldmath
Measurement of the $e^+e^-\to\pi^+\pi^-J/\psi$ cross section in the vicinity of 3.872 GeV
}

\author{
\begin{small}
\begin{center}
M.~Ablikim$^{1}$, M.~N.~Achasov$^{11,b}$, P.~Adlarson$^{70}$, M.~Albrecht$^{4}$, R.~Aliberti$^{31}$, A.~Amoroso$^{69A,69C}$, M.~R.~An$^{35}$, Q.~An$^{66,53}$, X.~H.~Bai$^{61}$, Y.~Bai$^{52}$
% ...
\\
\vspace{0.2cm}
(BESIII Collaboration)\\
\vspace{0.2cm} {\it
$^{1}$ Institute of High Energy Physics, Beijing 100049, People's Republic of China\\
% ...
$^{4}$ Bochum Ruhr-University, D-44780 Bochum, Germany\\
% ...
$^{11}$ G.I. Budker Institute of Nuclear Physics SB RAS (BINP), Novosibirsk 630090, Russia\\
% ...
$^{31}$ Johannes Gutenberg University of Mainz, Johann-Joachim-Becher-Weg 45, D-55099 Mainz, Germany\\
% ...
$^{35}$ Liaoning Normal University, Dalian 116029, People's Republic of China\\
% ...
$^{52}$ Southeast University, Nanjing 211100, People's Republic of China\\
$^{53}$ State Key Laboratory of Particle Detection and Electronics, Beijing 100049, Hefei 230026, People's Republic of China\\
% ...
$^{61}$ University of Jinan, Jinan 250022, People's Republic of China\\
% ...
$^{66}$ University of Science and Technology of China, Hefei 230026, People's Republic of China\\
% ...
$^{69}$ University of Turin and INFN, (A)University of Turin, I-10125, Turin, Italy; (B)University of Eastern Piedmont, I-15121, Alessandria, Italy; (C)INFN, I-10125, Turin, Italy\\
$^{70}$ Uppsala University, Box 516, SE-75120 Uppsala, Sweden\\
% ...
$^{b}$ Also at the Novosibirsk State University, Novosibirsk, 630090, Russia\\
% ...

}\end{center}

\vspace{0.4cm}
\end{small}
}

\begin{abstract}
We report a measurement of the cross section for the process $e^+e^-\to\pi^+\pi^-J/\psi$ around the $X(3872)$ mass in search for the direct formation of $e^+e^-\to X(3872)$ through the two-photon fusion process. No enhancement of the cross section is observed at the $X(3872)$ peak and an upper limit on the product of electronic width and branching fraction of $X(3872)\to\pi^+\pi^-J/\psi$ is determined to be $\Gamma_{ee}\times\mathcal{B}(X(3872)\to\pi^+\pi^-J/\psi)<7.5\times10^{-3}\,\text{eV}$ at $90\,\%$ confidence level under an assumption of total width of $1.19\pm0.21$ MeV. This is an improvement of a factor of about $17$ compared to the previous limit. Furthermore, using the latest result of $\mathcal{B}(X(3872)\to\pi^+\pi^-J/\psi)$, an upper limit on the electronic width $\Gamma_{ee}$ of $X(3872)$ is obtained to be $<0.32\,\text{eV}$ at the $90\,\%$ confidence level.

\end{abstract}


\maketitle

\oddsidemargin  -0.2cm
\evensidemargin -0.2cm


\section{Introduction}        

The observation of the $X(3872)$ state \footnote{also called $\chi_{c1}(3872)$, according to its quantum numbers} by the Belle collaboration in 2003~\cite{BelleObservation} and its confirmation by other experiments~\cite{XBabar,XCDF,XD0,XLHCb,XCMS,XBES} opened a new field of charmonium-like exotic states. Being difficult to fit into the conventional charmonium spectrum~\cite{PhysRevD.32.189,RyanReview}, the $XYZ$ states are candidates for tetraquarks, meson molecules, hybrid mesons, and kinematic effects~\cite{RevModPhys.90.015004,RyanReview,SoerenReview}. The $X(3872)$ state is probably the best known representative of these states. It has been observed in $B$ decays, in radiative transitions of the $Y(4260)$ resonance, as well as in inclusive $pp$ and $p\bar{p}$ collisions~\cite{PDG2020}. Up to now, its decays into six different final states are established~\cite{PDG2020,XtoPi0chic1,PhysRevD.100.094003}. Its mass is very close to the  threshold of $D\bar{D}^*$  production, which also represents one of its largest decay modes. This property leads naturally to the hypothesis that it is a meson molecule~\cite{XMolecule, RevModPhys.90.015004}. One important discriminant between different models is the $X(3872)$ width. Recently, two new measurements of its width were reported by the LHCb experiment~\cite{LHCb:2020xds,LHCb:2020fvo}, and an average of $1.19\pm0.21\,\text{MeV}$ based on these two new measurements is reported in Ref.~\cite{PDG2020}.

The quantum numbers $J^{PC}$ of the $X(3872)$ state are measured to be $1^{++}$~\cite{LHCbJPC}, which allows suppressed formation in $e^+e^-$ collisions via two-photon fusion~\cite{KUHN1979125,KAPLAN1978252}. Recently, a search for the $X(3872)$ in two-photon interactions in the process of $e^+e^-\to e^+e^- \pi^+\pi^- J/\psi$ was reported by the BELLE collaboration and evidence of $X(3872)$ production is found~\cite{PhysRevLett.126.122001}. Furthermore, the BESIII collaboration recently reported the first observation of the $1^{++}$ state $\chi_{c1}$ in direct $e^+e^-$ annihilation with a significance of 5.1$\sigma$~\cite{chic1_liutong}. These findings motivate the search for direct formation of $X(3872)$ via the two photon fusion process in $e^+e^-$ annihilation. Furthermore, knowledge of the electronic width $\Gamma_{ee}$ might help to reveal the nature of $X(3872)$. The current upper limit is $\Gamma_{ee}\times\mathcal{B}(X(3872)\to\pi^+\pi^-J/\psi)<0.13\,\text{eV}$ at the $90\,\%$ confidence level (C.L.), determined by BESIII via the initial-state radiation (ISR) process~\cite{BESIIILimit}. A theoretical prediction using vector meson dominance yields $\Gamma_{ee}\gtrsim0.03\,\text{eV}$ and a lower bound of $\Gamma_{ee}\times\mathcal{B}(X(3872)\to\pi^+\pi^-J/\psi)\gtrsim0.96\times10^{-3}\,\text{eV}$~\cite{TheoWidth}.

\begin{figure}
    \centering
    \includegraphics[width=0.45\textwidth]{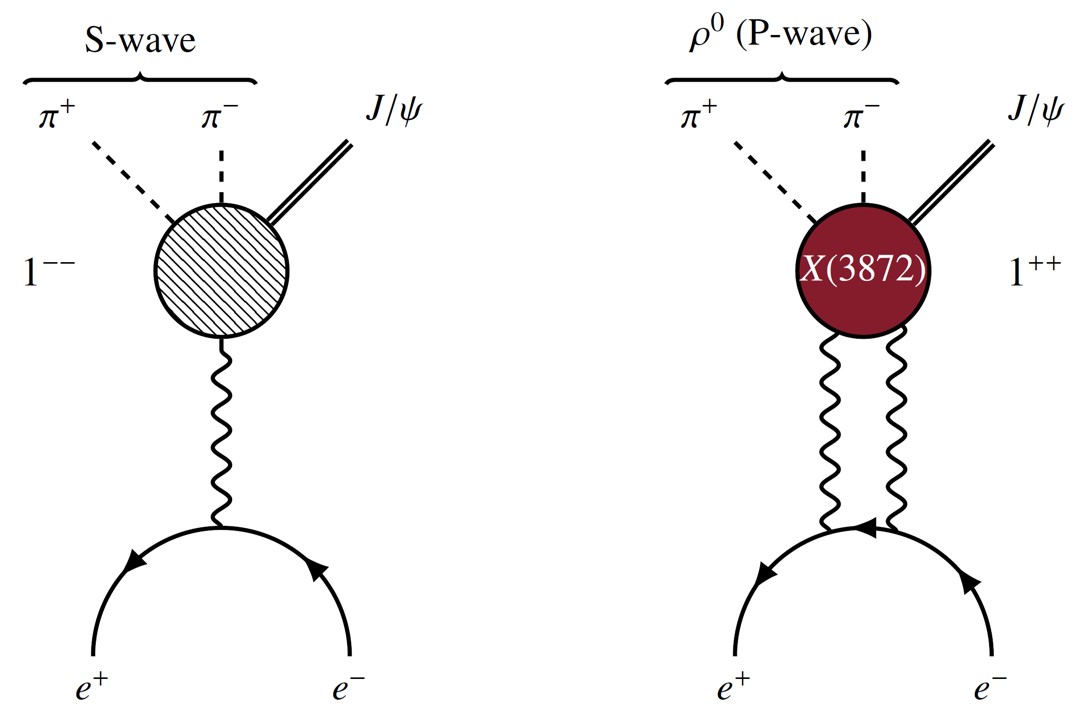}
    \caption{Feynman diagrams of the non-resonant continuum process $e^+e^-\to\pi^+\pi^-J/\psi$ (left) and the resonant signal process $e^+e^-\to X(3872)\to\pi^+\pi^-J/\psi$ (right), of which it is known that the $\pi^+\pi^-$ pair forms a $\rho^0$ meson~\cite{BelleLimit,XCMS,LHCbJPC}.}\label{fig:feynman}
\end{figure}

In this paper, the cross section $\sigma(e^+e^-\to\pi^+\pi^-J/\psi)$ around the $X(3872)$ mass is measured to search for the direct production of $X(3872)$ and measure the electronic width of $X(3872)$ using data sets collected by the BESIII detector. Two processes are considered, the non-resonant continuum process and the resonant signal process via two-photon fusion as shown in Fig.~\ref{fig:feynman}. An enhanced cross section is expected at the $X(3872)$ peak over the off-resonance region if the direct production process is significant.

\section{BESIII Detector and Data Samples}

The BESIII detector~\cite{ABLIKIM2010345} records symmetric $e^+e^-$ collisions provided by the BEPCII storage ring, which operates with a peak luminosity of $1\times10^{33}$~cm$^{-2}$s$^{-1}$  at the center-of-mass (c.m.) energy of $\sqrt{s}=3.89$~GeV. The cylindrical core of the BESIII detector covers 93\% of the full solid angle and consists of a helium-based multilayer drift chamber~(MDC), a plastic scintillator time-of-flight system~(TOF), and a CsI(Tl) electromagnetic calorimeter~(EMC), which are all enclosed in a superconducting solenoidal magnet providing a 1.0~T (0.9~T in 2012) magnetic field. The solenoid is supported by an octagonal flux-return yoke with resistive-plate-counter modules interleaved with steel, which are used for muon identification. The charged-particle momentum resolution at $1~{\rm GeV}/c$ is $0.5\%$, and the specific energy loss ($dE/dx$) resolution is $6\%$ for electrons from Bhabha scattering. The EMC measures photon energies with a resolution of $2.5\%$ ($5\%$) at $1$~GeV in the barrel (end-cap) region. The time resolution in the TOF barrel region is 68~ps, while that in the end-cap region is 110~ps. The end-cap TOF system was upgraded in 2015 using multi-gap resistive plate chambers, providing a time resolution of 60~ps~\cite{TOF1,TOF2,CAO2020163053}.

In this work, four data sets collected by the BESIII detector are used. Among the four data sets, two dedicated data sets were recorded in the vicinity of the $X(3872)$ mass, one directly on the $X(3872)$ peak with c.m.\ energy of 3871.3~MeV (on-resonance) and with a luminosity of 110.3~$\text{pb}^{-1}$, and the other one about 4~MeV below the peak at 3867.4~MeV (off-resonance) and with a luminosity of 108.9~$\text{pb}^{-1}$. For these two dedicated data sets, a Beam Energy Measurement System (BEMS)~\cite{Zhang_2016} provides the realtime measurement of the c.m.\ energy with an uncertainty smaller than 100 keV and of the energy spread with an uncertainty smaller than 200 keV. The resolution of c.m.\ energy from BEMS is smaller than the width of $X(3872)$, and far smaller than the 4 MeV energy difference between the on-resonance and off-resonance data sets. This ensures a negligible fraction of resonant contribution in the off-resonance data set at 3867.4 MeV. Two further off-resonance data sets at 3807.7 (50.5 $\text{pb}^{-1}$) and 3896.2 (52.6 $\text{pb}^{-1}$) MeV are used~\cite{lumi2013}. The luminosity of all data sets are determined by the analysis of large-angle Bhabha scattering events.

Simulated data samples produced with a {\sc geant4}-based~\cite{geant4} Monte Carlo (MC) package, which includes the geometric description of the BESIII detector and the
detector response, are used to determine detection efficiencies and to estimate backgrounds. The simulation models the beam energy spread and initial state radiation (ISR) in the $e^+e^-$
annihilations with the generator {\sc kkmc}~\cite{KKMC}. For each of the four different data samples, $5\times10^5$ signal MC events with $J/\psi\to e^+e^-$ and $5\times10^5$ signal MC events with $J/\psi\to\mu^+\mu^-$ are simulated with different models including VVPIPI and $\sigma$ PHSP~\cite{EvtGen}. The VVPIPI model describes the decay of a vector state to a vector state and $\pi^+\pi^-$ where the $\pi^+\pi^-$ system is dominated by a S-wave. The $\sigma$ PHSP is a model with phase space decay of $e^+e^-\rightarrow \sigma J/\psi$ and $\sigma \rightarrow \pi^+\pi^-$. To estimate the background contamination from non-$\pi^+\pi^-J/\psi$ events, background MC samples are generated, including $e^+e^-\to\gamma e^+e^-$, $\gamma\mu^+\mu^-$, $e^+e^-e^+e^-$, $e^+e^-\mu^+\mu^-$, $e^+e^-q\bar{q}$, $\pi^+\pi^-\pi^+\pi^-$, $K_S^0K^\pm\pi^\mp$, $K^+K^-\pi^+\pi^-$, and $e^+e^-\to\gamma_\text{ISR}\psi'\to\gamma_\text{ISR}\pi^+\pi^-J/\psi\to\gamma_\text{ISR}\pi^+\pi^-\ell^+\ell^-$. These backgrounds are scaled to the integrated luminosity of data for each energy point.

\section{Event Selection}

In the process of $e^+e^-\to\pi^+\pi^-J/\psi$, $J/\psi$ is reconstructed via its decay to lepton pairs $\ell^+\ell^-$ ($\ell=e,\,\mu$). The final state $\pi^+\pi^-\ell^+\ell^-$ consists of four charged tracks with zero net charge. The lepton tracks coming from the $J/\psi$ decay have larger momenta (in the lab frame) compared to the pion tracks. Tracks with  momenta lower than $0.6\,\text{GeV}/c$ are assumed to be pion candidates and lepton candidates are required to have momenta larger than $1.0\,\text{GeV}/c$. Each candidate event is required to have two pion candidates and two lepton candidates, both having opposite charges. The $J/\psi$ decay modes can be distinguished by the energy deposition in the EMC associated to the lepton tracks. Electrons deposit a large fraction of their energy, while the muons pass through the EMC leaving only a small energy deposition. The deposited energy in EMC is required to be smaller than 0.35 GeV for both muons and larger than 1.1 GeV for both electrons. Each event is required to have either two electron candidates or two muon candidates.

Candidate events of $e^+e^-\to\pi^+\pi^-\ell^+\ell^-$ are subjected to a kinematic fit with four constraints (4C) on the energy-momentum of the final states. The chi-square of kinematic fit, $\chi^2_{4C}$, is required to be smaller than 60. To remove photon conversion background, which could be misidentified as a $\pi^+\pi^-$ pair or $\pi^\pm e^\mp$, requirements of $\cos\theta_{\pi^+\pi^-} < 0.95$ and $\cos\theta_{\pi^\pm e^\mp}< 0.98$ are applied. Here, $\theta_{\pi^+\pi^-}$ is the opening angle between the two pion candidates, and $\theta_{\pi^\pm e^\mp}$ is the opening angle between the pion and oppositely charged electron candidates.

\begin{figure*}[htbp]
  \centering
  \subfigure{
  \label{gpp select etap}
  	\includegraphics[width=0.49\textwidth,page=5]{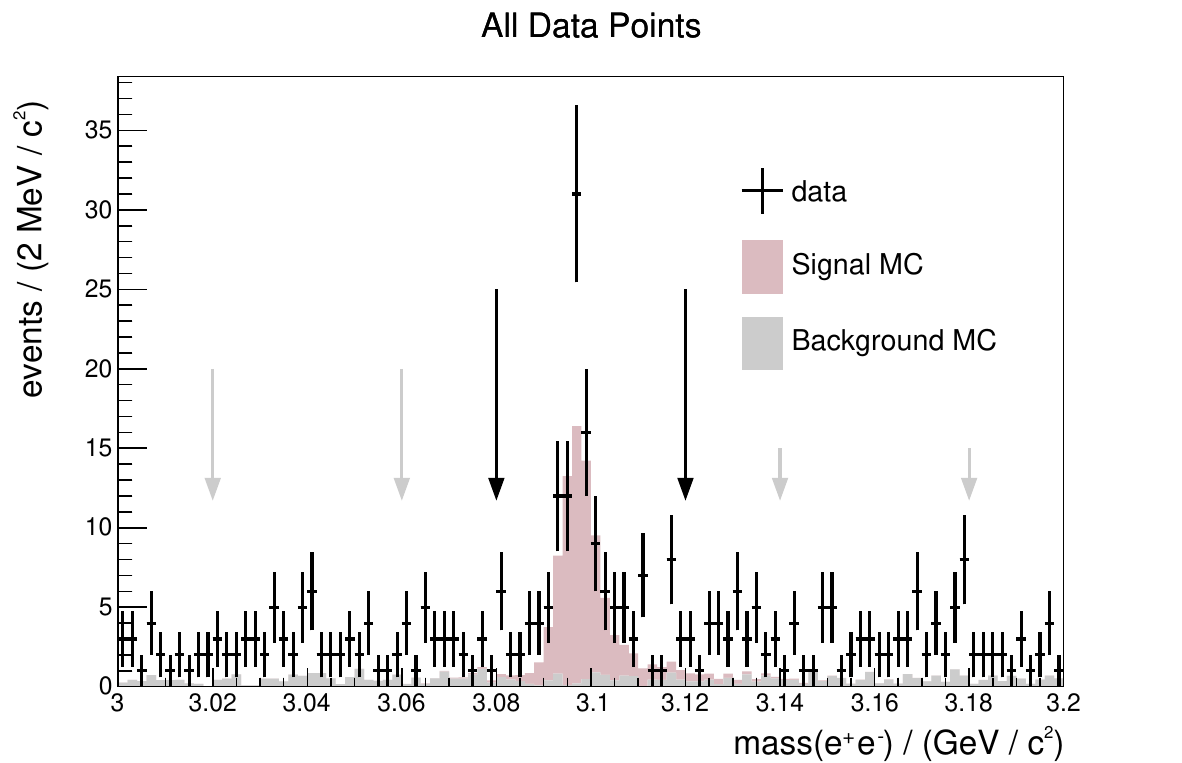}\put(-210,130){(a)}}%
  \subfigure{
  \label{epp select etap}
    \includegraphics[width=0.49\textwidth,page=10]{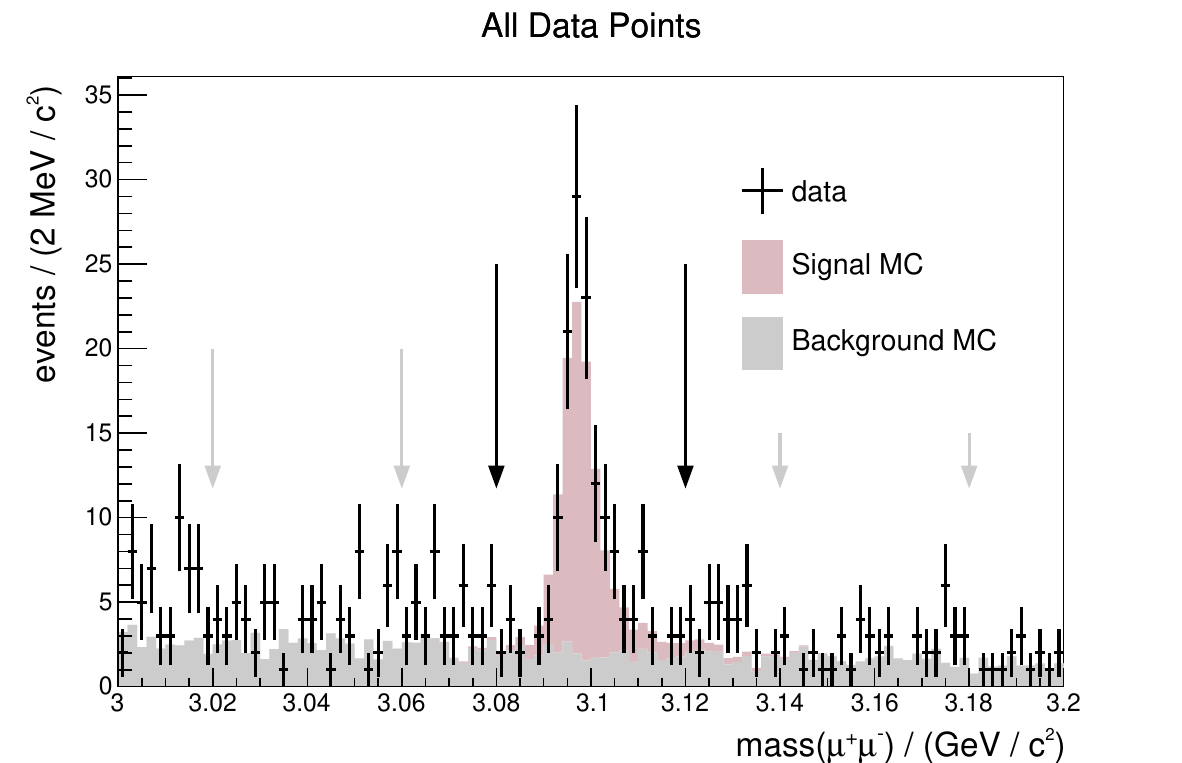}\put(-210,130){(b)}} %

   \caption{Comparison between data, signal MC, and background MC distributions of $m(\ell^+\ell^-)$. All four data sets are combined. The MC distributions are scaled to match the integrated luminosity of the data. The black (gray) arrows indicate the peak (sideband) region.}\label{fig:mll}
\end{figure*}

The invariant mass distributions of the di-lepton pairs that pass all selection criteria are shown in Fig.~\ref{fig:mll}. Here, the four data sets are combined together. The $J/\psi$ peak is clearly visible, and the peak region $m(\ell^+\ell^-)\in[3.08,3.12] \,\text{GeV}/c^2$ is indicated by a pair of black arrows as well as the sideband region $m(\ell^+\ell^-)\in[3.02,3.06]$ or $[3.14,3.18] \,\text{GeV}/c^2$ (gray arrows). In order to estimate the background contamination from non-$\pi^+\pi^-J/\psi$ events, the backgrounds $e^+e^-\to\gamma e^+e^-$, $\gamma\mu^+\mu^-$, $e^+e^-e^+e^-$, $e^+e^-\mu^+\mu^-$, $e^+e^-q\bar{q}$, $\pi^+\pi^-\pi^+\pi^-$, $K_S^0K^\pm\pi^\mp$, $K^+K^-\pi^+\pi^-$ were simulated. For the $e^+e^-$ mode, the background MC fails to describe the data. However, no peaking background is found from the simulation. This motivates an estimation of the background contamination by $J/\psi$ sidebands.

The number of signal events is determined from the $J/\psi$ peak. Instead of using the $\ell^+\ell^-$ invariant mass after the kinematic fit, the recoil mass of the two pions before the fit is used due to a better resolution. Independent unbinned maximum likelihood fits to the recoil mass of the two pions are performed for both $J/\psi$ decay modes in the range of $m(\ell^+\ell^-)\in[3.0, 3.2] \,\text{GeV}/c^2$. In the fit, the signal is described with a line shape derived from MC simulation, while the background is modeled as a linear function.

\section{Results}

The cross section of the process $e^+e^-\to\pi^+\pi^-J/\psi$ is determined using

\begin{align}
    \sigma=\frac{N_{\text{obs}}}{\int\mathcal{L}\,\text{d}t\cdot\epsilon\cdot(1+\delta)\cdot\mathcal{B}(J/\psi\to\ell^+\ell^-)}.\label{eq:xs}
\end{align}
Here, $N_{\text{obs}}$ is the number of observed signal events, $\int\mathcal{L}\,\text{d}t$ is the integrated luminosity, $\epsilon$ is the detection efficiency determined from MC samples, and $(1+\delta)$ is the radiative correction factor to account for ISR. It is calculated using the {\sc kkmc} event generator with the input line-shape updated iteratively until it converges.

\begin{figure}
    \centering
    \includegraphics[width=0.5\textwidth,page=1]{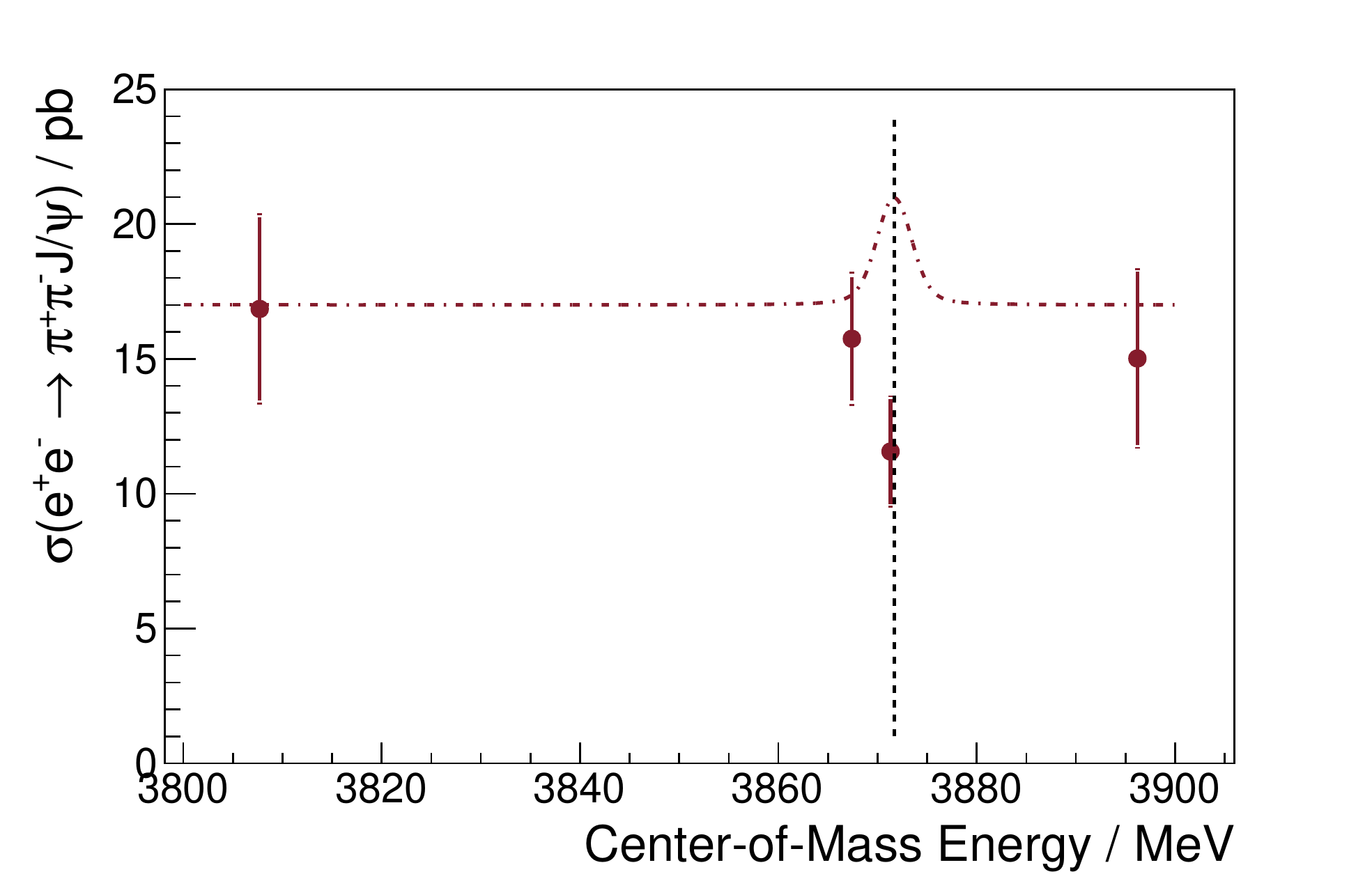}
    \caption{Cross section of $e^+e^-\to\pi^+\pi^-J/\psi$. The results of both $J/\psi$ decay modes are combined. The error bars represent the total uncertainties, {\it i.e.}, the quadratic sum of the statistical and systematic uncertainties. The dashed vertical line indicates the peak position of $X(3872)$. The dot-dashed curve shows an illustration of expected line-shape assuming $\Gamma_{ee}\times\mathcal{B}(X(3872)\to\pi^+\pi^-J/\psi) = 0.013$~eV.}\label{fig:xs}
\end{figure}

Table~\ref{tab:fit} shows the measured cross sections at the four energy points with two $J/\psi$ modes. The cross sections with two $J/\psi$ modes combined are also listed in Table~\ref{tab:fit} where the uncertainties are only statistical. In Fig.~\ref{fig:xs}, the cross sections are plotted with the systematic uncertainties which will be described later. In the plot, the dashed vertical line indicates the location of the $X(3872)$ peak. The dot-dashed curve shows the line-shape assuming a total width of 1.19~MeV and $\Gamma_{ee}\times\mathcal{B}(X(3872)\to\pi^+\pi^-J/\psi)$ = 0.013~eV. There is no interference between the resonant and continuum parts due to their different quantum numbers. The value of 0.013~eV is arbitrarily chosen as one order of magnitude lower than the previous upper limit~\cite{BESIIILimit}, and only for illustration purposes. From these data points, no enhancement of the cross section is observed at the $X(3872)$ peak. The cross section on-resonance is smaller than that of the off-resonance region, which can be observed as a dip in Fig.~\ref{fig:xs}. Given the low statistics, the dip is insignificant and consistent with a flat distribution within 1$\sigma$.

To interpret the cross section, the two processes in Fig.~\ref{fig:feynman} are considered. Due to the different quantum numbers of the final state in continuum process and the resonant $X(3872)$ formation, the total $e^+e^-\to\pi^+\pi^-J/\psi$ cross section is modeled as an incoherent sum of the two processes. The cross section from continuum is assumed to be a linear function of c.m.\ energy. The $X(3872)$ is modeled as a relativistic Breit-Wigner resonance. Only the $\pi^+\pi^-J/\psi$ decay mode is reconstructed in the cross section measurement. As a consequence, the corresponding branching fraction needs to be included in the line-shape parameterization:

\begin{align}
    \sigma(\sqrt{s})&=\sigma_{\text{cont}}+12\pi\,\frac{\Gamma_{\text{tot}}\Gamma_{ee}\times\mathcal{B}(X(3872)\to\pi^+\pi^-J/\psi)}{\left(s-m_0^2\right)^2+m_0^2\Gamma_{\text{tot}}^2},\label{eq:lineshape}
\end{align}
where $\sigma_{\text{cont}}$, $\Gamma_{\text{tot}}$, and $\Gamma_{ee}$ are the constant continuum cross section, the total width and the electronic width of the $X(3872)$, respectively. $m_0$ is the mass of $X(3872)$.

In 2020, two measurements on the total width were released as ($1.39\pm0.24\pm0.10$) MeV~\cite{LHCb:2020xds} and ($0.96^{+0.19}_{-0.18}\pm0.21$) MeV~\cite{LHCb:2020fvo}, resulting in an average value of ($1.19\pm0.21$) MeV~\cite{PDG2020}. In this analysis, we treat the product $\Gamma_{ee}\times\mathcal{B}(X(3872)\to\pi^+\pi^-J/\psi)$ as one parameter, and an upper limit on this product is measured. Then, we also calculate $\Gamma_{ee}$ using two different values for the branching fraction $\mathcal{B}(X(3872)\to\pi^+\pi^-J/\psi)$.  This was recently measured to be ($4.1\pm1.3$)\% by the BABAR collaboration~\cite{PhysRevLett.124.152001}, and the latest world average value is ($3.8\pm1.2$)\%~\cite{PDG2020}.

\begin{figure}
    \centering
    \includegraphics[width=0.5\textwidth]{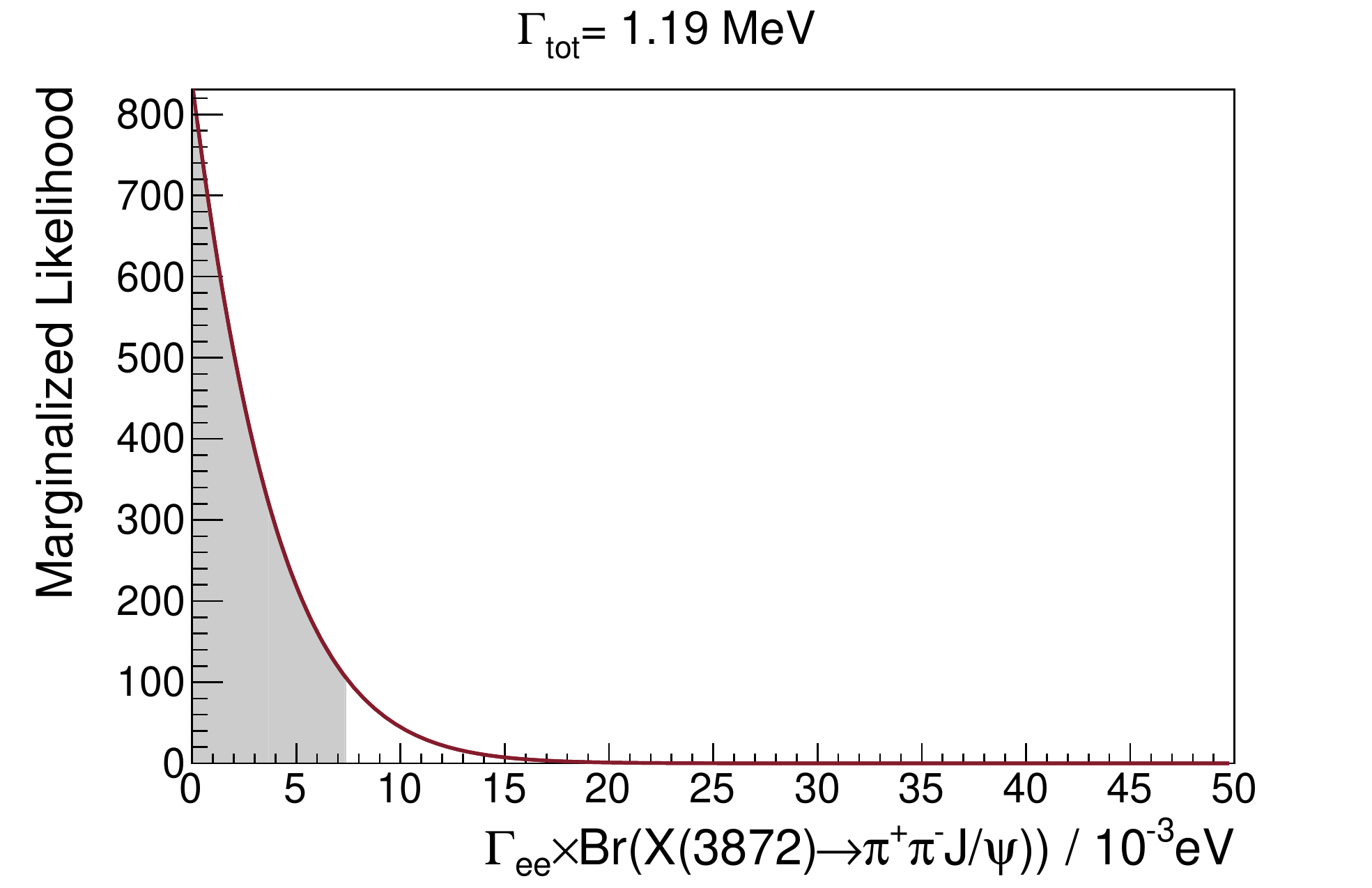}
    \caption{Determination of the upper limit on $\Gamma_{ee}\times\mathcal{B}$ for an assumed total width of $1.19\,\text{MeV}$. The gray area indicates the $90\,\%$ integral. The value of $\Gamma_{ee}\times\mathcal{B}$ at the right edge of that area is the upper limit on $\Gamma_{ee}\times\mathcal{B}$ at the $90\,\%$ confidence level.}\label{fig:limitfixGtot}
\end{figure}

\begin{figure}
    \centering
    \includegraphics[width=0.5\textwidth]{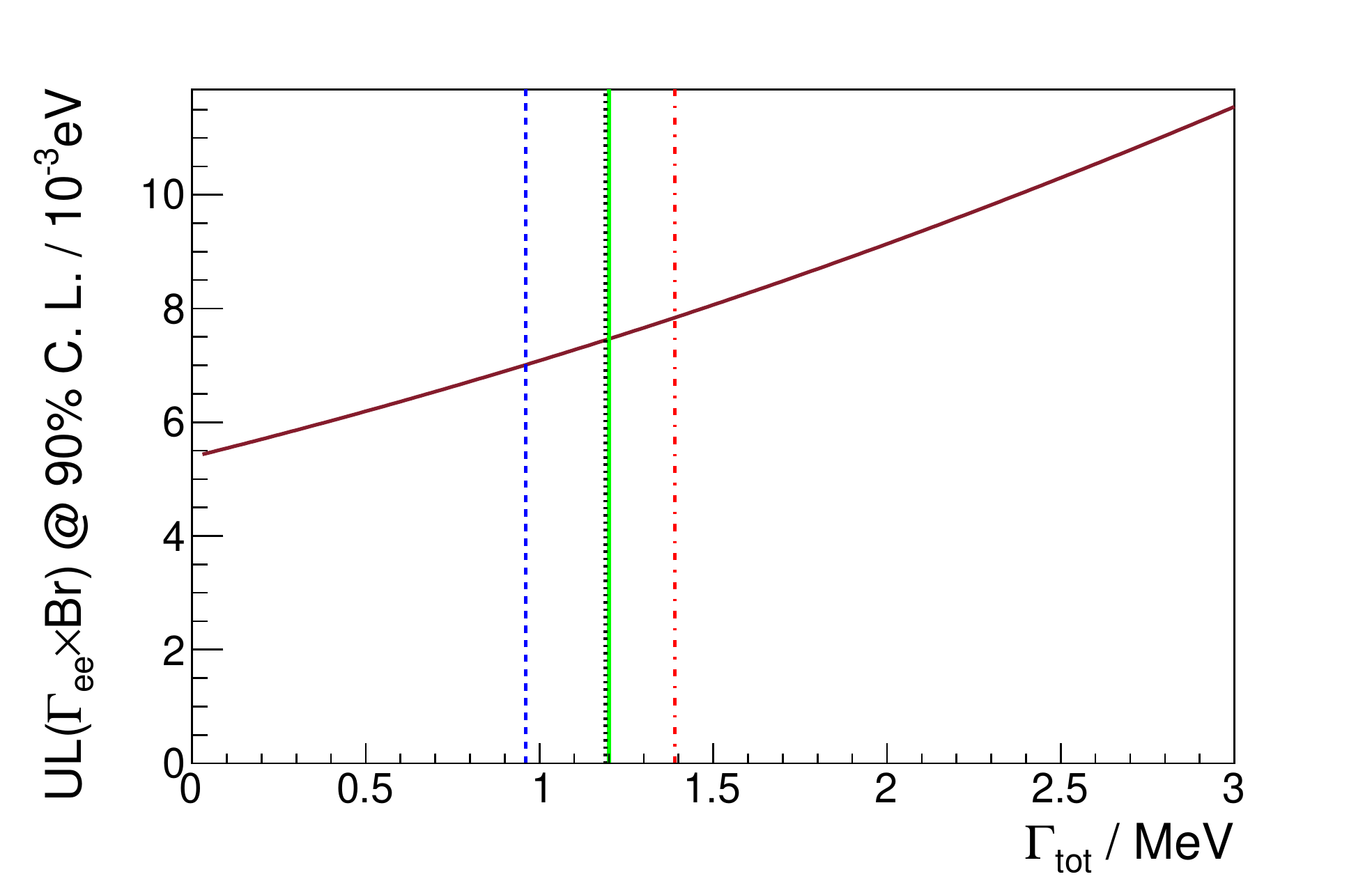}
    \caption{Upper limit on $\Gamma_{ee}\times\mathcal{B}$ at the $90\,\%$ confidence level as a function of the total width. The vertical lines indicates four different values of $\Gamma_{\rm tot}$, [0.96~\cite{LHCb:2020fvo}, 1.19~\cite{PDG2020}, $<$1.2~\cite{PDG2020}, 1.39~\cite{LHCb:2020xds}] MeV.}\label{fig:limitfuncGtot}
\end{figure}

In total, there are three unknown parameters left, $\sigma_{\text{cont}}$, $\Gamma_{\text{tot}}$, and $\Gamma_{ee}\times\mathcal{B}(X(3872)\to\pi^+\pi^-J/\psi)$ or $\Gamma_{ee}$. The mass $m_0$ is fixed to $(3871.65\pm0.06)\,\text{MeV}/c^2$~\cite{PDG2020}. Using the line-shape parameterization, an extended likelihood function depending on the cross section for each data set and $J/\psi$ mode is constructed. Upper limits for the electronic width are determined at different $\Gamma_{\text{tot}}$ inputs. Figure~\ref{fig:limitfixGtot} shows the determination of the upper limit on $\Gamma_{ee}\times\mathcal{B}$ for an assumed total width of $1.19\,\text{MeV}$. Figure~\ref{fig:limitfuncGtot} shows a scan of $\Gamma_{ee}\times\mathcal{B}$ in a wide range of $\Gamma_{\text{tot}}$ from 0 to 3 MeV. Four different $\Gamma_{\text{tot}}$, [0.96~\cite{LHCb:2020fvo}, 1.19~\cite{PDG2020}, $<$1.2~\cite{PDG2020}, 1.39~\cite{LHCb:2020xds}] MeV, are indicated with vertical lines. For each of the $\Gamma_{\text{tot}}$ inputs with a mean value and an uncertainty, a unique value of the upper limit of $\Gamma_{ee}\times\mathcal{B}$ is obtained by an integral of the likelihood over $\Gamma_{\text{tot}}$ based on a Gaussian assumption of $\Gamma_{\text{tot}}$. These upper limits are shown in Table~\ref{tab:GeeBr}, where both the statistical and systematic uncertainties have been considered.

\begin{table*}
    \centering
    \caption{Result of the fit to the dilepton mass distribution. Shown are the results of the two independent $J/\psi$ modes and a combined value. The uncertainties are only statistical. }\label{tab:fit}
    \begin{tabular}{lcccc}
	\hline\hline
	$\sqrt{s}\,/\,\text{MeV}$                                        & $3807.7 \pm 0.6$    & $3867.408 \pm 0.031$  & $3871.31 \pm 0.06$    & $3896.2 \pm 0.8$ \\
	$\int\mathcal{L}\,\text{d}t\,/\,\text{pb}^{-1}$                  & $50.5 \pm 0.5$      & $108.9 \pm 1.3$       & $110.3 \pm 0.8$       & $52.6 \pm 0.5$   \\
	$(1+\delta)$                                                     & $0.895$             & $0.895$               & $0.895$               & $0.895$          \\\hline
	$N_{obs}^{e^+e^-}$                                               & $19\pm5$            & $30\pm7$              & $24\pm6$              & $16\pm5$         \\
	$\epsilon^{e^+e^-}\,/\,\%$                                       & $31.78\pm0.08$      & $31.34\pm0.08$        & $31.29\pm0.08$        & $31.68\pm0.08$   \\
	$\sigma^{e^+e^-}\,/\,\text{pb}$                                  & $22.0 \pm 6.4$      & $16.4 \pm 3.6$        & $12.7 \pm 3.3$        & $17.0 \pm 5.2$    \\\hline
	$N_{obs}^{\mu^+\mu^-}$                                           & $18\pm5$            & $40\pm8$              & $29\pm6$              & $17\pm5$         \\
	$\epsilon^{\mu^+\mu^-}\,/\,\%$                                   & $45.38\pm0.10$      & $44.90\pm0.09$        & $44.72\pm0.09$        & $45.14\pm0.10$   \\
	$\sigma^{\mu^+\mu^-}\,/\,\text{pb}$                              & $14.8\pm4.0$        & $15.3\pm2.9$          & $10.9\pm2.4$          & $13.4\pm4.1$         \\\hline
	$\sigma^{\ell^+\ell^-}\,/\,\text{pb}$                            & $16.9\pm3.4$        & $15.7\pm2.3$          & $11.6\pm1.9$          & $15.0\pm3.2$     \\\hline
	\hline
    \end{tabular}

\end{table*}

\begin{table}
    \centering
    \caption{$\Gamma_{ee}$ and $\Gamma_{ee}\times\mathcal{B}(X(3872)\to\pi^+\pi^-J/\psi)$, using different values for the $X(3872)$ total width of ($0.96\pm0.21$) MeV~\cite{LHCb:2020fvo}, ($1.19\pm0.21$) MeV~\cite{PDG2020}, ($1.39\pm0.24$) MeV~\cite{LHCb:2020xds}, and the current upper limit of 1.2 MeV~\cite{PDG2020}. Two input values for $\mathcal{B}(X(3872)\to\pi^+\pi^-J/\psi)$ are used, $\mathcal{B}_1 = (4.1\pm1.3)\%$~\cite{PhysRevLett.124.152001}, and $\mathcal{B}_2 = (3.8\pm1.2)\%$.}\label{tab:GeeBr}
    \begin{tabular}{lrrr}
        \hline\hline
        & \multicolumn{2}{c}{$\Gamma_{ee}\,/10^{-3}\,\text{eV}$} & $\Gamma_{ee}\times\mathcal{B}\,/10^{-3}\,\text{eV}$ \\
        $\Gamma_{\text{tot}}\,/\,\text{MeV}$   &  $\mathcal{B}_1$    & $\mathcal{B}_2$    &---   \\
        \hline
        $0.96\pm0.21$                     & $<293$ & $<305$ & $<7.0$ \\
        $1.19\pm0.21$                     & $<309$ & $<322$ & $<7.5$ \\
        $1.39\pm0.24$                     & $<324$ & $<338$ & $<7.9$ \\
        $<1.2$                            & $<270$ & $<282$ & $<6.5$ \\
        \hline\hline
    \end{tabular}

\end{table}

\section{Systematic Uncertainties}

There are two different kinds of systematic uncertainties that have to be treated separately. Uncertainties of the first kind affect the cross section measurement; they comprise the uncertainty of the integrated luminosity, the tracking efficiency, the branching fraction of $J/\psi$ decays to lepton pairs, the line-shape of the continuum process, the kinematic fit, the decay model, the fit to the invariant mass of $J/\psi$, and the resolution of the invariant mass of lepton pair.

The integrated luminosity is determined using the same strategy as in Ref.~\cite{lumi2013}. The uncertainty of the tracking efficiency is 1\% per track yielding a 4\% uncertainty~\cite{PhysRevD.100.051101}. The uncertainty from the line-shape of the continuum process is estimated by changing the assumed flat distribution to be the line-shape reported in Ref.~\cite{ZhiqingPaper}. In the kinematic fits, the helix parameters of charged tracks are corrected to reduce the discrepancy between data and MC simulation as described in Ref.~\cite{helixparam}. The difference between MC samples with and without the correction is taken as the uncertainty from the kinematic fits. The uncertainty associated with the decay model is estimated by the efficiency difference between $\sigma$ PHSP model and VVPIPI model~\cite{EvtGen}. In the fit to the $m(\ell^+\ell^-)$ distribution, the background is modeled as a linear function. To estimate the uncertainty from the background modeling in the fit, an alternative fit using a quadratic function is performed to $1000$ pseudo data sets. These pseudo data sets are sampled from the real data using a bootstrap method, where the events from the real data are allowed to be picked multiple times. The fit results are averaged and the difference to the nominal fit is taken as systematic uncertainty associated with the fit. In order to account for the difference of the resolution of $m(\ell^+\ell^-)$ in data and simulation, an alternative fit to the $m(\ell^+\ell^-)$ spectrum is performed. The signal probability density function is convolved with a Gaussian and the resulting difference in $\sigma(e^+e^-\to\pi^+\pi^-J/\psi)$ is less than 0.1\% and can be neglected. These systematic uncertainties are summarized in Table~\ref{tab:sysErrxsX}.

\begin{table*}
    \centering
    \caption{Relative systematic uncertainties (in $\%$) on the measured cross section $\sigma(e^+e^-\to\pi^+\pi^-J/\psi)$. The total uncertainty is the quadratic sum of the individual uncertainties.}\label{tab:sysErrxsX}
    \begin{tabular}{lcccccccc}
        \hline\hline

    & \multicolumn{2}{c}{$3807.7\,\text{MeV}$}    & \multicolumn{2}{c}{$3867.4\,\text{MeV}$}  & \multicolumn{2}{c}{$3871.3\,\text{MeV}$}    & \multicolumn{2}{c}{$3896.2\,\text{MeV}$} \\
	Source & $e^+e^-$ & $\mu^+\mu^-$ & $e^+e^-$ & $\mu^+\mu^-$ & $e^+e^-$ & $\mu^+\mu^-$ & $e^+e^-$ & $\mu^+\mu^-$\\
	\hline
	$\int\mathcal{L}\,\text{d}t$ & \multicolumn{2}{c}{$1.0$} & \multicolumn{2}{c}{$1.2$} & \multicolumn{2}{c}{$0.7$} & \multicolumn{2}{c}{$1.0$} \\
	Tracking & \multicolumn{2}{c}{$4.0$} & \multicolumn{2}{c}{$4.0$} & \multicolumn{2}{c}{$4.0$} & \multicolumn{2}{c}{$4.0$} \\
	$\mathcal{B}(J/\psi\to \ell^+\ell^-)$ & $0.5$ & $0.6$ & $0.5$ & $0.6$ & $0.5$ & $0.6$ & $0.5$ & $0.6$ \\
	Line-shape                            & $0.8$ & $0.8$ & $1.2$ & $1.7$ & $1.3$ & $1.3$ & $0.7$ & $0.6$ \\
	Kinematic fit                         & $0.9$ & $0.7$ & $0.8$ & $0.7$ & $0.9$ & $0.7$ & $0.9$ & $0.7$ \\
	Decay model                           & $2.1$ & $3.6$ & $2.7$ & $4.0$ & $2.2$ & $3.7$ & $2.4$ & $4.0$ \\
	$m(\ell^+\ell^-)$ fit                 & $4.9$ & $1.2$ & $2.2$ & $1.7$ & $4.2$ & $7.7$ & $11.8$& $4.2$ \\
	\hline
	Total                                 & $6.8$ & $5.7$ & $5.7$ & $6.3$ & $6.5$ & $9.6$ & $12.8$& $7.2$ \\
	\hline\hline
    \end{tabular}

\end{table*}

Uncertainties of the second kind affect the line-shape parameterization. They include the uncertainties associated with the $X(3872)$ mass, $\sqrt{s}$, and the beam energy spread. The values for these parameters are sampled from the corresponding distributions to calculate the likelihood. With a subsequent integral over the likelihood, these uncertainties have already been included in the electronic width measurement.

\section{Summary}

In summary, the cross sections of the process $e^+e^-\to\pi^+\pi^-J/\psi$ are measured at four different c.m.\ energies around the $X(3872)$ mass. The results of these measurements are listed in Table~\ref{tab:fit}. Since the direct production of the $X(3872)$ is not observed, an upper limit on $\Gamma_{ee}\times\mathcal{B}$ is determined. For an assumed total width of $1.19\pm0.21\,\text{MeV}$, the upper limit on $\Gamma_{ee}\times\mathcal{B}$ is determined to be $7.5\times10^{-3}\,\text{eV}$ at the $90\,\%$ C.L., with an improvement of a factor of about $17$ compared to the previous limit~\cite{BESIIILimit}. The upper limits on $\Gamma_{ee}\times\mathcal{B}$ are also calculated for other values of $\Gamma_{\text{tot}}$, as summarized in Table~\ref{tab:GeeBr}. There is no conflict to the theoretical prediction of $\Gamma_{ee}\times\mathcal{B}\gtrsim0.96\times10^{-3}\,\text{eV}$~\cite{TheoWidth}.
Using the recently released measurements of $\mathcal{B}(X(3872)\to\pi^+\pi^-J/\psi)$, an upper limit on $\Gamma_{ee}$ is reported for the first time.

\section*{\boldmath ACKNOWLEDGMENTS}
The BESIII collaboration thanks the staff of BEPCII and the IHEP computing center for their strong support. This work is supported in part by National Natural Science Foundation of China (NSFC) under Contracts Nos. 11975278, 11635010, 11735014, 11835012, 11935015, 11935016, 11935018, 11961141012, 12022510, 12025502, 12035009, 12035013, 12192260, 12192261, 12192262, 12192263, 12192264, 12192265; National Key Research and Development Program of China under Contracts Nos. 2020YFA0406300, 2020YFA0406400;  the Chinese Academy of Sciences (CAS) Large-Scale Scientific Facility Program; Joint Large-Scale Scientific Facility Funds of the NSFC and CAS under Contracts Nos. U2032105, U2032109, U1832207; CAS Key Research Program of Frontier Sciences under Contract No. QYZDJ-SSW-SLH040; 100 Talents Program of CAS; CAS Interdisciplinary Innovation Team; Key Research Program of the Chinese Academy of Sciences under Contract No. XDB34030301; INPAC and Shanghai Key Laboratory for Particle Physics and Cosmology; ERC under Contract No. 758462; European Union's Horizon 2020 research and innovation programme under Marie Sklodowska-Curie grant agreement under Contract No. 894790; German Research Foundation DFG under Contracts Nos. 443159800, Collaborative Research Center CRC 1044, GRK 2149; Istituto Nazionale di Fisica Nucleare, Italy; Ministry of Development of Turkey under Contract No. DPT2006K-120470; National Science and Technology fund; National Science Research and Innovation Fund (NSRF) via the Program Management Unit for Human Resources and Institutional Development, Research and Innovation under Contract No. B16F640076; STFC (United Kingdom); Suranaree University of Technology (SUT), Thailand Science Research and Innovation (TSRI), and National Science Research and Innovation Fund (NSRF) under Contract No. 160355; The Royal Society, UK under Contracts Nos. DH140054, DH160214; The Swedish Research Council; U. S. Department of Energy under Contract No. DE-FG02-05ER41374.

\bibliography{bib_directX}{}

\end{document}